\documentclass[twocolumn,showpacs,preprintnumbers, nofootinbib,prd,aps,reprint]{revtex4-1}
\usepackage{hyperref}
\usepackage{longtable}
\usepackage[caption = false]{subfig}
\usepackage{graphicx}
\usepackage{dcolumn}
\usepackage{bm}
\usepackage{amsmath,amssymb,amsfonts}
\usepackage{latexsym}
\usepackage{bbm}
\usepackage{url} 
\usepackage{soul}
\usepackage{multirow}
\usepackage{braket}
\usepackage{hyperref} 
\usepackage{slashed}

\begin{document}
\preprint{LA-UR-25-23006}

\title{Towards the determination of CP-odd pion-nucleon couplings}

\author{Shohini Bhattacharya$^{a,b}$}
\email{shohinib@uconn.edu}
\author{Kaori Fuyuto$^{b}$}
\email{kfuyuto@lanl.gov}
\author{Emanuele Mereghetti$^{b}$}
\email{emereghetti@lanl.gov}
\author{Thomas R.~Richardson$^{c,d}$}
\email{thomas.richardson@berkeley.edu}
\affiliation{$^{a}$ Department of Physics, University of Connecticut, Storrs, CT 06269, U.S.A.}
\affiliation{$^{b}$ Theoretical Division, Los Alamos National Laboratory, Los Alamos, NM 87545, USA}
\affiliation{$^{c}$ Department of Physics, University of California, Berkeley, CA 94720, USA}
\affiliation{$^{d}$Nuclear Science Division, Lawrence Berkeley National Laboratory, Berkeley, CA 94720, USA}

\bigskip

\date{\today}

\begin{abstract}

The nucleon matrix elements (NMEs) associated with quark chromo-magnetic dipole moments (cMDMs) play a crucial role in determining the CP-odd pion-nucleon couplings induced by quark chromo-electric dipole moments. In recent years, it has been argued that the NMEs of cMDMs can be related to the third moment of the nucleon’s higher-twist (specifically, twist-three) parton distribution function (PDF) $e(x)$, which can, in principle, be measured through dihadron production in semi-inclusive deep inelastic scattering processes.
By applying the spin-flavor expansion to the cMDM operators in the large-$N_c$ limit, where $N_c$ is the number of quark colors, we show that the NMEs receive contributions not only from the twist-three PDF $e(x)$ but also from an additional, previously neglected nucleon form factor. 
Incorporating constraints from the spin-flavor expansion, recent experimental data on $e(x)$, as well as model calculations of $e(x)$, we estimate the NMEs of the cMDM operators. Our analysis indicates that the NMEs are dominated by the nucleon form factors, and the cMDM contributions to pion-nucleon couplings can be comparable to those from the quark sigma terms.

\end{abstract}

\maketitle

\section{Introduction}
One of the major mysteries that the Standard Model (SM) cannot account for is the matter-antimatter asymmetry in our Universe. 
Interactions that distinguish between particles and antiparticles, and thus imply the violation of CP symmetry, are essential ingredients for creating the asymmetry. Although the SM contains a CP-violating phase, known as the Cabibbo-Kobayashi-Maskawa phase, it turns out that the CP violation is not large enough to produce the observed asymmetry \cite{Gavela:1993ts, Gavela:1994dt, Huet:1994jb, Konstandin:2003dx}. Thus, the solution to the mystery requires  physics beyond the Standard Model (BSM) to have some new sources of CP violation.

Searches for permanent electric dipole moments (EDMs)  offer the best opportunity to reveal new sources of CP violation. Since the SM predictions are too small to be observed in current and next-generation EDM experiments, the observation of any EDMs points to the existence of new CP violation. To date, EDM searches have been conducted utilizing several species, e.g., nucleons, atoms, and molecules, enabling us to scrutinize the underlying CP nature.
The sensitivity of molecular EDM  experiments that probe the electron EDM has remarkably improved in recent years. The latest limit is $|d_e|<4.1\times 10^{-30}~e~$cm \cite{Roussy:2022cmp}, which now severely constrains BSM scenarios that correlate with the electron EDM \cite{Fuyuto:2015ida, Fuyuto:2017ewj, Fuyuto:2019svr, Bodeker:2020ghk}. 
On the other hand, neutron and atomic EDMs can be induced even if CP-violating interactions  predominantly affect the quark sector. Their experimental bounds, for example, $|d_n|<1.8\times 10^{-26}~e~$cm for neutron \cite{Abel:2020pzs} and $|d_{\rm Xe}|<1.4\times 10^{-27}~e~$cm \cite{Sachdeva:2019rkt} for $^{129}$Xe, also play a significant role in probing various BSM models, and next-generation searches aim to improve this sensitivity by a factor of 10 \cite{Ito:2017ywc, n2EDM:2021yah, TUCAN:2018vmr, 120XeEDM_LANL}.

Along with the experimental efforts, it is an urgent issue on the theory side to provide quantitative connections between quark-level CP-violating interactions and the EDMs of nucleons, nuclei and atoms. 
For the nucleon and atomic EDMs, the challenge first arises from the nonperturbative nature of Quantum Chromodynamics (QCD) at low energy, and the extraction of nucleon matrix elements (NMEs) of CP-violating (CPV) operators. 
Lattice QCD offers a first-principle method for the calculation of such NMEs, and various lattice groups have reported the results of the nucleon EDMs induced by the QCD $\theta$ term \cite{Guo:2015tla, Abramczyk:2017oxr, Dragos:2019oxn, Alexandrou:2020mds, Bhattacharya:2021lol, Liang:2023jfj} and by the quark EDMs \cite{Gupta:2018lvp}. 
In the case of the quark chromo EDMs (cEDMs), 
to most important NMEs are the nucleon EDMs and two CP-odd pion-nucleon interactions, which give the dominant contribution to the nuclear Schiff moment of diamagnetic atoms.
At the moment, there exist preliminary Lattice QCD calculations of the nucleon EDM  
\cite{Abramczyk:2017oxr,Bhattacharya:2023qwf},
while no Lattice QCD estimates of the pion-nucleon couplings induced by the cEDM is available.
An alternative approach has been explored in Ref. \cite{Seng:2018wwp}; the study relates the NMEs of chromo-magnetic dipole moments (cMDMs), which are essential inputs for CPV pion-nucleon couplings induced by the quark cEDMs, to the nucleon's twist-3 parton distribution function (PDF) denoted by $e(x)$. By using a handful of the PDF experimental data \cite{Efremov:2002ut}, Ref. \cite{Seng:2018wwp} extracted the NMEs of the cMDMs.

In this paper, following the approach in \cite{Seng:2018wwp}, we estimate the NMEs of the cMDMs by imposing constraints from the spin-flavor symmetry in the large-$N_c$ limit of QCD and incorporating the latest experimental results and model calculations of the twist-3 PDF. 
The appropriate analysis of the spin-flavor expansion of the cMDM operators is in tension with a crucial assumption in Ref.~\cite{Seng:2018wwp} in which one of the nucleon form factors in the NMEs is estimated to be negligible. 
Together with the fit to the twist-3 PDF data recently updated by the CLAS Collaboration as well as the model calculations, we estimate the NMEs of the cMDMs and show that they receive a dominant contribution from the nucleon form factor. We finally compare the NMEs with those from the quark sigma terms and
discuss their contributions to the CP-odd pion-nucleon couplings.

The paper is organized as follows. In section \ref{sec:pncoupling}, we introduce the NMEs relevant for the CP-odd pion-nucleon couplings induced by the quark cEDMs. In section \ref{sec:twist3}, we review the twist-3 PDF and its correlations with the NMEs of the cMDMs. Section \ref{sec:spin-flavor-symmetry} is dedicated to the application of the spin-flavor symmetry to the cMDMs. In section \ref{sec:results}, we present the fitting results of the NMEs taking into account the constraints from the spin-flavor symmetry and argue the contribution to the pion-nucleon couplings. Finally, we conclude in section \ref{sec:conclusion}.

\section{CP-odd pion-nucleon couplings}\label{sec:pncoupling}

We start by writing down the QCD $\theta$ term and the quark cEDM operators:
\begin{align}
{\cal L}_{\rm CPV}\supset -\frac{\bar{\theta}}{32\pi^2}G^A_{\mu\nu}\tilde{G}^{A\mu\nu}-i\frac{g_s}{2}\bar{q}(\sigma\cdot G)\tilde{d}_{\rm cEDM}\gamma_5q,
\end{align}
where 
$g_s$ denotes the strong coupling constant,
$\tilde{G}^A_{\mu\nu}=(1/2)\epsilon_{\mu\nu\alpha\beta}G^{A\alpha\beta}$, $\sigma\cdot G=\sigma^{\mu\nu}G^A_{\mu\nu}T^A$, $q=(u,d)^T$, and $\tilde{d}_{\rm cEDM}={\rm diag}(\tilde{d}_u,\tilde{d}_d)$. 
With a $U(1)_A$ transformation of the quark fields, we can move the QCD $\theta$ term into the CPV quark mass term. Then performing vacuum alignment provides \cite{Dashen:1970et, Baluni:1978rf, Crewther:1979pi, deVries:2012ab, Bsaisou:2014oka}
\begin{align}
{\cal L}_{\rm CPV}\supset&~\bar{q}\left[m_*\left(\bar{\theta}-\bar{\theta}_{\rm ind}\right)+r\tilde{d}_{\rm cEDM}\right]i\gamma_5q\nonumber\\
&-i\frac{g_s}{2}\bar{q}(\sigma\cdot G)\tilde{d}_{\rm cEDM}\gamma_5q.
\end{align}
The reduced quark mass $m_*$ is given by
\begin{align}
m_*=\left(\frac{1}{m_u}+\frac{1}{m_d}\right)^{-1}=\frac{\bar{m}(1-\varepsilon^2)}{2},
\end{align}
with $\bar{m}=(m_u+m_d)/2$ and $\varepsilon=(m_d-m_u)/2\bar{m}$. 
The induced theta term $\theta_{\rm ind}$ is described by the vacuum condensate ratio $r$ 
\begin{align}
\bar{\theta}_{\rm ind}=r\left(\frac{\tilde{d}_u}{m_u}+\frac{\tilde{d}_d}{m_d}\right),
\hspace{0.3cm}
r=\frac{1}{2}\frac{\langle 0|\bar{q}g_s\sigma^{\mu\nu}G_{\mu\nu}q|0\rangle}{\langle 0|\bar{q}q|0\rangle}.
\end{align}
The above expression indicates that, if there are other CPV interactions than the QCD theta, $\bar{\theta}$ relaxes to $\bar{\theta}_{\rm ind}$ in the Peccei–Quinn mechanism. The value of the vacuum condensate ratio is roughly estimated by a QCD Sum Rule from which $r=(0.4\pm0.05)$ GeV$^2$ \cite{Ioffe1982,Kogan1984, Narison:2007spa}.

The quark cEDMs become a source of CPV pion-nucleon interactions. Following the derivation in \cite{deVries:2012ab}  based on $SU(2)$ chiral perturbation theory, one can describe the hadronic interactions as 
\begin{align}
    \label{eq:cp-violating-chiral}
{\cal L}_{\pi N}=-\frac{\bar{g}_0}{2F_{\pi}}\bar{N}\tau\cdot \pi N-\frac{\bar{g}_1}{2F_{\pi}}\pi^0\bar{N}N,
\end{align}
with the nucleon field $N=(p,n)^T$ and the pion decay constant $F_{\pi}=92.2~$MeV. The isovector and isoscalar couplings, respectively, are given by
\begin{align}
\bar{g}_0&=\frac{1}{2}\left(\tilde{d}_u+\tilde{d}_d\right)\left(\sigma^3_C+\frac{r\sigma^3}{\bar{m}\varepsilon} \right), \label{eq:pN_0} \\
\bar{g}_1&=-\left(\tilde{d}_u-\tilde{d}_d\right)\left(\sigma^0_C-\frac{r\sigma^0}{\bar{m}} \right), \label{eq:pN_1}
\end{align}
where the nucleon matrix elements, i.e., the sigma and cMDM terms are introduced as
\begin{align}
\sigma^0&=\frac{\bar{m}}{2m_N}\langle P|\bar{q}q|P\rangle,\\
\sigma^3&=\frac{\bar{m}\varepsilon}{m_N}\langle P|\bar{q}\tau^3q|P\rangle,\\
\sigma^0_C&=\frac{1}{4m_N}\langle P|\bar{q}g_s(\sigma\cdot G) q|P\rangle, \label{eq:cMDM-isoscalar}\\
\sigma^3_C&=-\frac{1}{2m_N}\langle P|\bar{q}g_s(\sigma\cdot G)\tau^3q|P\rangle, \label{eq:cMDM-isovector}
\end{align}
with a proton state $|P\rangle$ and the nucleon mass $m_N$. 
Eqs. \eqref{eq:pN_0}
and \eqref{eq:pN_1} are a consequence of the
spontaneously broken chiral symmetry of the QCD Lagrangian, which allows to relate nucleon matrix elements of isoscalar (isovector) CP-even operators to isovector (isoscalar) CP-odd pion-nucleon couplings 
\cite{Crewther:1979pi}.
For the nucleon sigma terms, the FLAG 2024 results are \cite{FlavourLatticeAveragingGroupFLAG:2024oxs},  
\begin{align}
    \sigma^0&=
    \begin{cases}
    60.9~{\rm MeV}, & (N_f=2+1+1),\\
    42.2~{\rm MeV}, & (N_f=2+1),
    \end{cases}\\
    \hspace{0.5cm}
    \sigma^3&=2.74~{\rm MeV},
\end{align}
where the isovector sigma term is obtained by multiplying $g_S^{u-d}$ by $\bar{m}\epsilon$ with $g_S^{u-d}=1.085$ \cite{FlavourLatticeAveragingGroupFLAG:2024oxs}\footnote{In the case of the isovector scalar charge, $2+1-$flavor calculations present the consistent results $g_S^{u-d}=1.083$ evaluated at 2~GeV\cite{FlavourLatticeAveragingGroupFLAG:2024oxs}.}, $\bar{m}=3.49$ MeV, $m_d=4.7$ MeV, and $m_u/m_d=0.462$ evaluated at 2~GeV \cite{ParticleDataGroup:2024cfk}. Note that the above sigma terms are scale independent.

On the other hand, the study of the cMDM terms is still ongoing \cite{Walker-Loud}, and no lattice QCD results are available yet. Because of this, 
the best estimates of pion nucleon couplings still rely on QCD sum rule calculations \cite{Pospelov:2001ys,Lebedev:2004va,Pospelov:2005pr,Fuyuto:2012yf}.
The determination of the NMEs is essential to provide accurate predictions of EDMs and investigate the possibility of CPV sources in BSM physics. 
In the next section, we will review the discussion held in \cite{Seng:2018wwp}, presenting how the cMDM terms can be related to the nucleon twist-three chiral-odd distribution functions $e^q(x)$.

\section{Nucleon twist-three parton distribution function}
\label{sec:twist3}
Our goal is to estimate the NMEs of the cMDMs, denoted as $\sigma^{0/3}_C$, which correlate with the third moment of the twist-three (chiral-odd) PDF $e^q(x)$. 
In this section, we will carefully outline the derivation and highlight the key caveats involved in applying the relations and assumptions employed in Ref.~\cite{Seng:2018wwp}. 
Before formally defining this PDF, we first provide a brief overview of twist classification in the context of PDFs within hadron structure physics.

By virtue of asymptotic freedom in QCD, the cross-section of high-energy processes can be factorized into a ``hard part" (perturbative) and a ``soft part" (non-perturbative).
For instance, the cross-section for inclusive unpolarized Deep Inelastic Scattering (DIS) is given by:
\begin{equation}
    \sigma_{\text{DIS}}(x, Q^2) = \sum_i \left[ H_{\text{DIS}}^i \otimes f_i \right] (x, Q^2),
    \label{e:obs}
\end{equation}
where the convolution integral is defined as
\begin{equation}
    [a \otimes b](x) \equiv \int_x^1 \frac{d\xi}{\xi} a\left( \frac{x}{\xi} \right) b(\xi).
\end{equation}
Here, the index $i$ runs over all parton species, including quarks, antiquarks, and gluons. The variable $x$ denotes the Bjorken scaling parameter, while $Q^2$ represents the energy scale of the hard interaction. The function $H_{\text{DIS}}$ describes the hard-scattering kernel, which is process-dependent and calculable using perturbative QCD. Meanwhile, $f_i$ encapsulates the non-perturbative aspects of the cross section, encoding the internal structure of hadrons. This non-perturbative component of the cross section can be systematically expanded in powers of the large energy scale, $Q$, that characterizes the process under consideration. For example, the function $f_i$ in Eq.~(\ref{e:obs}) can be expressed in a twist expansion, where the twist is defined as the mass dimension minus the spin:
\begin{equation}
    f_i = f_i^{(0)} + \frac{f_i^{(1)}}{Q} + \frac{f_i^{(2)}}{Q^2} + \dots \, .
\end{equation}
Here, $f_i^{(0)}$ denotes the leading-twist (twist-2) term, while $f_i^{(1)}$ and $f_i^{(2)}$ correspond to twist-3 and twist-4 terms, respectively. The leading-twist term has a clear probabilistic interpretation, representing the number density of partons carrying a given momentum fraction in the infinite momentum frame. In contrast, higher-twist terms involve correlations between multiple partons and introduce power-suppressed effects that encode multipartonic dynamics. Since these effects are power suppressed in observables, they are notoriously difficult to measure. Consequently, most studies focus primarily on leading-twist PDFs, as they are the most accessible. Nevertheless, higher-twist PDFs, though less extensively explored, have gained increasing attention, particularly in the context of lattice calculations~\cite{Bhattacharya:2020cen,Bhattacharya:2020xlt,Bhattacharya:2020jfj}.

We now proceed to define and examine the fundamental properties of $e^q(x)$, which have been extensively analyzed in Ref.~\cite{Jaffe:1991kp, Jaffe:1991ra, Efremov:2002qh} and related studies. For completeness, we will quote key equations from the reference \cite{Efremov:2002qh} and include additional details to establish its connection with the cMDM.
To begin, consider the definition of $e^q(x)$ in terms of a quark bilinear operator, expressed as a matrix element within a proton state $|P\rangle$ \cite{Jaffe:1991kp, Jaffe:1991ra}:
\begin{equation}
e^q(x) = \frac{1}{2m_N} \int \frac{d\lambda}{2\pi} e^{i\lambda x} \langle P | \bar{\psi}^q(0) [0, \lambda n] \psi^q (\lambda n) | P \rangle \, ,
\end{equation}
where $\psi^q$ is a quark field of flavor $q$.
Here, $n^\mu$ is a basis vector on the light cone and $[0, \lambda n]$ denotes the Wilson line, 
\begin{align}
    [0,\lambda n] \equiv {\cal P}\exp\left[ig_s \int^{\lambda}_0d\tau~n^\mu G^A_{\mu} (\tau) T^A \right] .
\end{align}
The Wilson line describes the parallel transport of a color charge along the lightlike direction $n^\mu$, ensuring the gauge invariance of nonlocal quark bilinear operators in PDFs. 
In this context,   the path-ordering operator $\mathcal{P}$ ensures that the non-Abelian gauge fields are correctly ordered along the integration path.
The support of $e(x)$ lies within $-1 \leq x \leq 1$ and satisfies the symmetry relation  $e^q(-x) = e^{\overline{q}}(x)$.

\begin{figure}[h]
    \centering
    \includegraphics[width=0.4\textwidth]{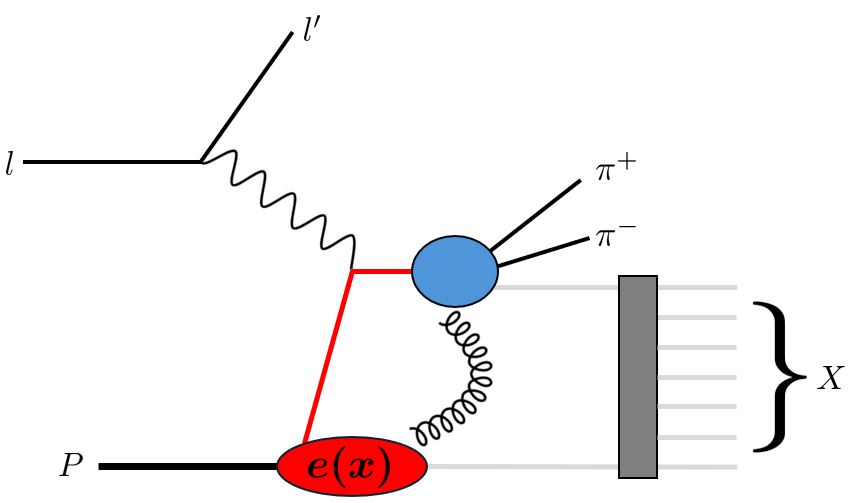} 
    \caption{Semi-Inclusive Deep Inelastic Scattering (SIDIS) process, $l + P \rightarrow l' + \pi^+ + \pi^- + X$, where $l$ represents the incoming lepton, $P$ is the proton, and $X$ denotes the unobserved hadronic final state. The red blob corresponds to the twist-3 distribution $e(x)$, while the hadronization of the struck quark produces the detected $\pi^+$ and $\pi^-$ mesons.}
    \label{fig:sidis}
\end{figure}

The scalar density operator $\bar{\psi}(0) [0, \lambda n] \psi(\lambda n)$ can be decomposed using the QCD equation of motion, resulting in its expression as a sum of distinct components—specifically, the ``singular" term~\footnote{Note that the Jaffe-Ji sum rule connects the first moment of the flavor-singlet combination of $e(x)$ to the sigma term $\sigma^0$~\cite{Jaffe:1991kp}.}, ``pure twist-three" term (related to quark-gluon correlations), and ``quark mass" term (related to the current quark mass):
\begin{equation}
e^q(x) = e^q_{\text{sing.}}(x) + e^q_{\text{tw3}}(x) + e^q_{\text{mass}}(x),
\end{equation}
where:
\begin{align}
e^q_{\text{sing.}}(x) &  = \frac{1}{2m_N} \int \frac{d\lambda}{2\pi} e^{i\lambda x} \langle P | \bar{\psi}^q(0) \psi^q(0) | P \rangle \nonumber \\[0.2cm]
& = \delta(x) \frac{1}{2m_N} \langle P | \bar{\psi}^q(0) \psi^q(0) | P \rangle \, , \label{e:sing}\\[0.4cm]
e^q_{\text{tw3}}(x) & = \frac{1}{4m_N} \int \frac{d\lambda}{2\pi} e^{i\lambda x} \lambda^2 \mathcal{F}^q_{\text{tw3}}(\lambda) \, , \nonumber \\[0.2cm]
\mathcal{F}^q_{\text{tw3}}(\lambda) & = \int_{0}^{1} du \int_{0}^{u} dv \, \langle P | \bar{\psi}^q(0) \sigma^{\alpha\beta} n_{\beta} [0, v\lambda n] \nonumber \\[0.2cm]
& \times g_s G_{\alpha\nu}(v\lambda n) n^{\nu} [v\lambda n, u\lambda n] \psi^q(u\lambda n) | P \rangle \, , \\[0.4cm]
e^q_{\text{mass}}(x) & = -\frac{m_q}{m_N} \int \frac{d\lambda}{4\pi} e^{i\lambda x} i\lambda \nonumber \\[0.2cm]
& \times \int_{0}^{1} du \, \langle P | \bar{\psi}^q(0) \slashed{n} [0, u\lambda n] \psi^q(u\lambda n) | P \rangle \, ,
\end{align}
where $m_q$ denotes the quark mass. The study of $e(x)$ has been a topic of debate for over 30 years due to its intriguing theoretical properties. In particular, one interesting yet sometimes controversial aspect of $e(x)$ is the potential existence of delta-function singularities, $\delta(x)$ (see Eq.~(\ref{e:sing})), and their implications for certain sum rules, specifically the moments of $e(x)$. For a comprehensive reference that compiles existing model calculations of $e(x)$ in the literature, as well as a discussion on which models predict the appearance of delta-function singularities, see Ref.~\cite{Bhattacharya:2020jfj}. 
Regarding the processes sensitive to this PDF, Ref.~\cite{Jaffe:1991kp} reported that $e(x)$ can be accessed in the unpolarized Drell-Yan process, though only at twist-4, where it mixes with unknown quark-gluon-quark correlations. Later, it was shown that $e(x)$ convoluted with the twist-2 Collins fragmentation function can be probed through a single-spin asymmetry in semi-inclusive deep inelastic scattering (SIDIS) processes of longitudinally polarized leptons off unpolarized protons~\cite{Levelt:1994np}. 
A particularly clean channel to access $e(x)$ is di-hadron production in SIDIS~\cite{Bacchetta:2003vn} (see Fig.~\ref{fig:sidis}), for which the CLAS collaboration recently extracted the relevant single-spin asymmetry~\cite{Hayward:2021psm}. With determinations of $e(x)$ now available~\cite{Courtoy:2022kca}, we will later use these results to assess its contribution to the cMDM.

We now present a compilation of model-independent moment relations for $e^q(x)$ with a particular focus on the third moment, which plays a crucial role in establishing a connection with cMDM. 
For convenience, we define the $n$-th moment of the function $q(x)$ as follows:
\begin{equation}
{\cal M}_n [q] \equiv \int^1_{-1} dx x^{n-1} q(x) \, \quad n \geq 1 \, .
\end{equation}
This integral extends over the full range of $x$, ensuring that both the quark and antiquark contributions are included. 
The third moment of $e^q(x)$, which arises from both the pure twist-three term and the quark mass term, can be written as:
\begin{equation}
{\cal M}_3 [e^q] = {\cal M}_3 [e^q_{\rm tw3}] + {\cal M}_3 [e^q_{\rm mass}] \, ,
\label{e:sum}
\end{equation}
where~\footnote{The components of a light-cone four-vector $a^\mu$ are defined as $a^\pm = (1/\sqrt{2})(a^0 \pm a^3)$.}:
\begin{align}
{\cal M}_3 [e^q_{\rm tw3}] & = \frac{1}{4 m_N (P^+)^2} \nonumber \\
& \times \sum_{i=1}^{2} \langle P | \bar{\psi}^q(0) g_s \sigma^{+ i} G^{+ i}(0) \psi^q(0) | P \rangle \, ,\\[0.2cm]
\mathcal{M}_3[e_{\text{mass}}^q] & = \frac{m_q}{m_N} \mathcal{M}_2[f_1^q] \, ,
\end{align}
where $f^q_1$ is the unpolarized twist-2 PDF. The third moment of \( e^q(x) \) is primarily determined by the pure twist-three contribution rather than the sum of this term with the quark mass contribution. Using PDF inputs from LHAPDF~\cite{Buckley:2014ana}, we find that the quark mass term is approximately \( \sim 10^{-4} \), making its effect relatively small. Consequently, we assume that ${\cal M}_3 [e^q_{\rm tw3}]$ dominates in Eq.~(\ref{e:sum}), leading to the approximation:  
\begin{equation}
{\cal M}_3 [e^q] \approx {\cal M}_3 [e^q_{\rm tw3}] \, 
\label{e:approx}.
\end{equation}

We now aim to establish a connection between the third moment of ${\cal M}_3 [e^q]$ and the cMDM sigma terms. As mentioned in Ref.~\cite{Seng:2018wwp}, this can be demonstrated by analyzing the parameterization of the spin-averaged matrix element involving two free Lorentz indices from \( \bar{\psi} G \cdot \sigma \psi \):
\begin{align}
\langle P | \bar{\psi}^q(0) g_s G^{\alpha \mu}(0) \sigma_{\alpha}^{\ \nu} \psi^q(0) | P \rangle & = A^q m_N \left( m_N^2 g^{\mu \nu} - P^\mu P^\nu \right) \nonumber \\[0.2cm]
& + B^q m_N P^\mu P^\nu,
\label{eq:para}
\end{align}
where \( A^q \) and \( B^q \) are dimensionless and Lorentz-invariant but scale-dependent nucleon form factors. The connection between the cMDM sigma term and $A^q$ and $B^q$ becomes evident upon observing that:
\begin{align}
\langle P | \bar{\psi}^q(0) g_s G^{\alpha \mu}(0) \sigma_{\alpha \mu} \psi^q(0) | P \rangle = \left(3 A^q + B^q\right) m_N^3,
\label{e:e1}
\end{align}
therefore
\begin{align}
\sigma^0_C&=\frac{1}{4 m_N} \left( 
    \langle P \vert \bar{u} g_s \sigma_{\mu \nu} G^{\mu \nu} u \vert P \rangle 
    + 
    \langle P \vert \bar{d} g_s \sigma_{\mu \nu} G^{\mu \nu} d \vert P \rangle 
\right) \nonumber \\[0.2cm]
&=\frac{1}{4}m^2_N\left[3\left(A^u+A^d\right)+B^u+B^d\right], \label{eq:sigma0C}\\[0.2cm]
\sigma^3_C&=-\frac{1}{2 m_N} \left( 
    \langle P \vert \bar{u} g_s \sigma_{\mu \nu} G^{\mu \nu} u \vert P \rangle 
    - 
    \langle P \vert \bar{d} g_s \sigma_{\mu \nu} G^{\mu \nu} d \vert P \rangle 
\right) \nonumber \\[0.2cm]
&=-\frac{1}{2}m^2_N\left[3\left(A^u-A^d\right)+B^u-B^d\right] \label{eq:sigma3C}.
\end{align}
On the other hand, it is straightforward to check that ${\cal M}_3 [e^q_{\rm tw3}]$ is determined by a different combination of \( A^q \) and \( B^q \):
\begin{align}
{\cal M}_3 [e^q_{\rm tw3}] = \dfrac{A^q - B^q}{4}.
\label{e:e2}
\end{align}
We now use Eq.~(\ref{e:e2}) to express $A^q$ in terms of $B^q$ and substitute it into Eqs.~(\ref{eq:sigma0C}) and (\ref{eq:sigma3C}) to obtain:
\begin{align}
\sigma_C^{0}
& \approx m_N^2 \left( 3~( {\cal M}_3 [e^u] + {\cal M}_3 [e^d] ) + B^u + B^d \right) \label{e:main1}\, ,\\[0.3cm]
\sigma_C^{3}
& \approx -2 m_N^2 \left( 3~( {\cal M}_3 [e^u] - {\cal M}_3 [e^d] ) + B^u - B^d \right).
\label{e:main2}
\end{align}
Here, the approximation in Eq.~(\ref{e:approx}) was applied to the above equations.
This result highlights the central message of Ref.~\cite{Seng:2018wwp}: the cMDM sigma terms are intrinsically connected to the third moment of \( e^{u/d}(x) \), namely \( {\cal M}_3 [e^{u/d}] \), or equivalently, to the third moment of its pure twist-three component, \( {\cal M}_3 [e^{u/d}_{\rm tw3}] \), as well as to two unknown nucleon form factors, \( B^{u,d} \).

Now it should be emphasized that the previous study neglected the contributions from the unknown form factors $B^{u/d}$ to the cMDMs and concluded that the quark sigma terms, $\sigma^{0,3}$, dominate the pion-nucleon couplings \cite{Seng:2018wwp}. 
The assumption of $B^q=0$ does not originate from analyzing the cMDM operator itself, rather this assumption is based on an analysis of the quark bilinear $\bar{q}\sigma_{\alpha}^{~\nu}q$ in the non-relativistic limit.
Specifically, $\bar{q}\sigma_{\alpha}^{~\nu}q$ is nonzero only when $\alpha,\nu\neq 0$, which would suggest that the right-hand side of Eq.~\eqref{eq:para} must vanish when $\mu=\nu=0$ implying that $B^q = 0$.
Ref.~\cite{Hatta:2020ltd} previously cautioned against neglecting $B^q$ based on a twist analysis, although its primary focus was on the Weinberg operator. 
The operator $\bar{\psi} \sigma^{+i} G^{+i} \psi \sim (A - B)/4$ is of twist-3, while $\bar{\psi} \sigma \cdot G \psi \sim 3 A m_N^3 + B m_N^3$ is of twist-5. 
Since these matrix elements involve different linear combinations of $A$ and $B$, setting $B = 0$ would incorrectly imply their proportionality, despite belonging to different twist sectors~\footnote{We thank Y. Hatta for sharing this perspective with us.}. 
In the next section, we apply the spin-flavor expansion \cite{Dashen:1994qi} to the cMDM operator, and we will demonstrate that the form factors can take nonzero values that potentially provide a dominant contribution to the cMDMs.


\section{Spin-flavor symmetry}
    \label{sec:spin-flavor-symmetry}

In the large-$N_c$ limit, where $N_c$ is the number of colors, combinatoric arguments suggest that meson-baryon scattering amplitudes should be $O(N_c^0)$ \cite{Witten:1979kh}.
This scaling along with unitarity implies that there is an infinite number of degenerate baryons that fill out irreducible representations of SU$(2 N_F)$ \cite{Dashen:1993as, Dashen:1993jt, Dashen:1994qi, Gervais:1983wq, Gervais:1984rc, Carone:1993dz}, where $N_F$ is the number of active quark flavors; here, we take $N_F = 2$.
In particular, this means that the nucleon and the $\Delta$ resonance are degenerate states at leading order.

At large but finite $N_c$, the baryon matrix elements of any QCD operator containing $m$ quark bilinears can be expanded in terms of the SU(4) generators as \cite{Dashen:1994qi}
    \begin{equation}
            \label{eq:spin-flavor-expansion}
        \mathcal O^{(m)}_{QCD} = N_c^m  \sum_{n,s,t} c_n  \left( \frac{\hat J^i}{N_c} \right)^{s} \left( \frac{\hat I^a}{N_c} \right)^{t} \left( \frac{\hat G^{jb}}{N_c} \right)^{n - s - t} \, ,
    \end{equation}
where it should be understood that we are considering the baryon matrix elements $\langle B'| \mathcal O^{(m)}_{QCD} |B \rangle$ and $| B \rangle$ is a generic baryon state.
The spin-flavor generators are
    \begin{align}
        \hat J^i & = q^\dagger \frac{\sigma^i}{2} q \, , \\
        \hat I^a & = q^\dagger \frac{\tau^a}{2} q \, , \\
        \hat G^{ia} & = q^\dagger \frac{\sigma^i \tau^a}{4} q \, .
    \end{align}
The matrices $\sigma^i$ ($\tau^a$) are the usual Pauli matrices in spin (isospin) space.
This is the same basis used in the nonrelativistic quark model; however, this expansion does not assume the validity of the quark model even though the spin-flavor expansion and the quark model are equivalent at leading order.
Physical baryon states have $O(N_c^0)$ spin and isospin matrix elements whereas the combined spin-flavor generator can have $O(N_c)$ matrix elements, so we have the scaling
    \begin{equation}
        \begin{split}
        \langle B' | \hat J^i | B \rangle, \, \langle B' | \hat I^a | B \rangle & \sim O(N_c^0) \, , \\
        \langle B' | \hat G^{ia} | B \rangle & \sim O(N_c) \, .
        \end{split}
    \end{equation}

The parity and time-reversal transformation properties as well as the spin and isospin indices of each term on the right hand side of Eq.~\eqref{eq:spin-flavor-expansion} need to match those of the QCD operator on the left hand side.
Furthermore, operator reduction rules allow us to eliminate several redundant terms at a given order of $n$-body operators in Eq.~\eqref{eq:spin-flavor-expansion}.
Also, the series can be truncated for $N_c = 3$ at the level of three-body operators, i.e. $n = 3$.
Previously, this technique has been used to analyze baryon masses \cite{Jenkins:1995td}, magnetic moments \cite{Jenkins:1994md}, and axial couplings \cite{Dashen:1994qi} (see Ref.~\cite{Jenkins:1998wy} for a review).
These operators generally consist of a single quark bilinear.
Here, we construct the appropriate spin-flavor expansions of the cMDM operators in Eqs.~\eqref{eq:cMDM-isoscalar} and \eqref{eq:cMDM-isovector} and the parameterization in Eq.~\eqref{eq:para} \textit{including} the gluon field strength tensor, which is, to the best of our knowledge, the first direct application of this expansion to these types of operators.

The spin-flavor expansion of the isoscalar cMDM is similar to the expansion of the baryon mass operator.
This operator is even under both parity and time-reversal transformations and has no free Lorentz indices.
Therefore, the structure of each term in the spin-flavor expansion must reproduce these properties.
The proton matrix element is thus expanded as
    \begin{equation}
        \begin{split}
            \langle P | \bar q \sigma^{\mu \nu} G_{\mu \nu} q | P \rangle  & = \bra{P} m_0^{(u+d)} N_c + m_2^{(u+d)} \frac{J^2}{N_c} \ket{P} \\
            & = m_0^{(u+d)} N_c +  \frac{3}{4 N_c} m_2^{(u+d)} \, ,
        \end{split}
        \label{eq:cMDM-isoscalar-sf}
    \end{equation}
where $m_0^{(u+d)}$ and $m_2^{(u+d)}$ are undetermined coefficients that are at most $O(N_c^0)$.
In the second line we have used the fact that the physical proton state has spin 1/2 and $\langle P | J^2 | P \rangle = \frac{3}{4}$.
However, the term proportional to $J^2$ will only be different when comparing nucleon matrix elements to $\Delta$ resonance matrix elements.
The matrix elements of this operator between neutron states will be identical to those of the proton.

A similar expansion is performed for the isovector cMDM operator.
However, we must retain a free isospin index,
    \begin{equation} 
        \begin{split}
            \langle P | \bar q \sigma^{\mu \nu} G_{\mu \nu} \tau^3 q | P \rangle  & = m_1^{(u-d)} I^3 + \cdots \\
            & = \frac{m_1^{(u-d)}}{2} + \cdots \, ,
        \end{split}
        \label{eq:cMDM-isovector-sf}
    \end{equation}
where $m_1^{(u-d)}$ is again an $O(N_c^0)$ undetermined coefficient, and the dots represent higher order terms that we have omitted.
We have used $\langle P | I^3 | P \rangle = \frac{1}{2}$ in the second line for the physical proton state.
The neutron matrix elements would come with the opposite sign; however, the isovector operator is suppressed by a factor of $1/N_c$ relative to the isoscalar operator.

We can also obtain the scaling of the individual $u$ and $d$ contributions to the matrix elements by taking the appropriate linear combinations of Eqs.~\eqref{eq:cMDM-isoscalar-sf} and \eqref{eq:cMDM-isovector-sf}.
Keeping the leading contributions leads to
    \begin{align}
        \langle P | \bar u \sigma^{\mu \nu} G_{\mu \nu} u | P \rangle  & = \frac{1}{2} m_0^{(u+d)} N_c \left[ 1 + O(1/N_c) \right] \, , \\
        \langle P | \bar d \sigma^{\mu \nu} G_{\mu \nu} d | P \rangle  & = \frac{1}{2} m_0^{(u+d)} N_c \left[ 1 + O(1/N_c) \right] \, .
    \end{align}
Therefore, the $u$ and $d$ matrix elements between proton states are equal to one another up to $O(1/N_c)$ corrections.

These scalings can then be mapped to the low energy constants or couplings of the chiral Lagrangian.
In particular, we have $\bar g_0 \sim O(N_c^0)$ and $\bar g_1 \sim O(N_c)$ if we omit the scaling of $g_s$ and $m_N$ in the definitions of the sigma terms.
Including the scaling of $F_\pi \sim O(\sqrt{N_c})$, the coefficients in Eq.~\eqref{eq:cp-violating-chiral} are in agreement with those of Ref.~\cite{Samart:2016ufg} obtained from an analysis of the parity and time-reversal-invariance violating two-nucleon potential, which provides an independent validation of the analysis presented here.

Let us now consider the parameterization 
in Eq.~(\ref{eq:para}).
Now, we need to examine the spin-flavor expansions of the various combinations of $\mu$ and $\nu$ separately.
Additionally, we stress that we are not considering the structure of the quark bilinear by itself as considered in Ref.~\cite{Seng:2018wwp}.
Rather, we are considering the expansion of the complete operator containing the gluon field strength.
We will work in the rest frame, $P^\mu = (m_N, 0)$.

First, take $\mu = \nu = 0$.
The first term in Eq.~\eqref{eq:para} will vanish and the second term will have an expansion identical to Eq.~\eqref{eq:cMDM-isoscalar-sf} for the isoscalar combination and Eq.~\eqref{eq:cMDM-isovector-sf} for the isovector combination only with different expansion coefficients,
    \begin{align}
        B^{(u+d)} m_N^3   & = t_{B,0}^{(u+d)} N_c + t_{B,2}^{(u+d)} \frac{J^2}{N_c} \, , \label{eq:B-isoscalar-sf} \\
        B^{(u-d)} m_N^3  & = t_{B,1}^{(u-d)} I^3  \label{eq:B-isovector-sf} \, .
    \end{align}

When $\mu = i$ and $\nu = j$ in the rest frame, only the first term in Eq.~\eqref{eq:para} survives.
The leading terms in the spin-flavor expansion are
    \begin{align}
        A^{(u+d)} m_N^3 g^{ij} & = N_c t_A^{(u+d)} \left\{ \frac{G^{ia}}{N_c}, \frac{G^{ja}}{N_c} \right\} \, , \\
        A^{(u-d)} m_N^3 g^{ij} & = N_c t_A^{(u-d)} \left\{ \frac{G^{i3}}{N_c}, \frac{J^j}{N_c} \right\} \, .
    \end{align}
Again, the matrix element of the isoscalar combination gives the dominant contribution and is $O(N_c)$ while the isovector combination is $1/N_c$ suppressed.
Moreover, $A^{(u \pm d)}$ and $B^{(u \pm d)}$ scale in the same way.
In particular,
    \begin{align}
        \left| A^{(u+d)} \right| \sim \left| B^{(u+d)} \right| \sim O(N_c) \, , \\
        \left| A^{(u-d)} \right| \sim \left| B^{(u-d)} \right| \sim O(N_c^0) \, ,
    \end{align}
such that the $u$ and $d$ contributions to the matrix elements are equal to one another up to $\sim 30\%$ corrections.

\begin{figure*}[ht!]
    \subfloat[\label{fig:figure1}]{\includegraphics[width=0.5\textwidth]{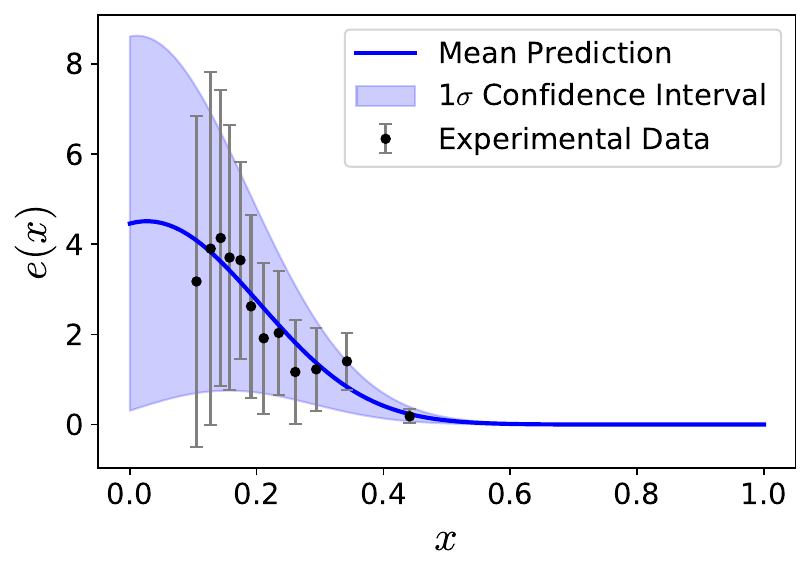}}
    \hfill
    \subfloat[\label{fig:figure2}]{\includegraphics[width=0.5\textwidth]{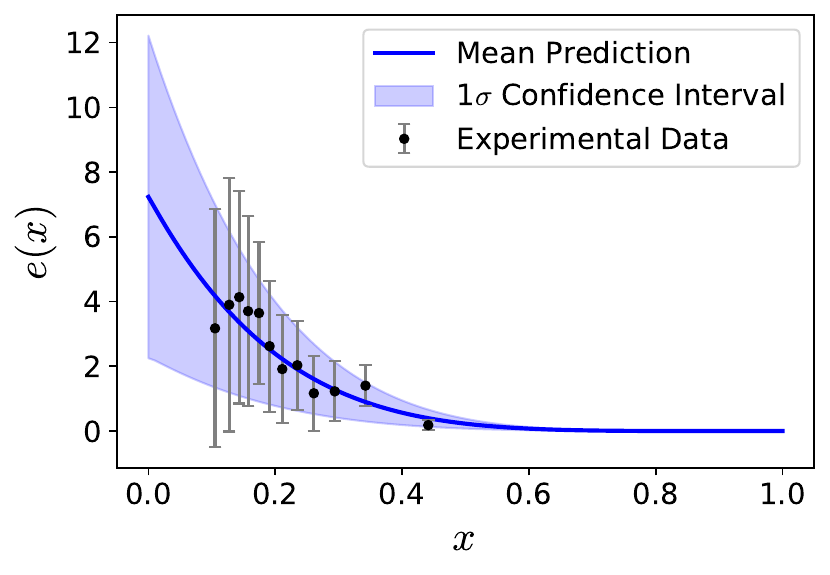}}
    \caption{(a) A Gaussian fit to the CLAS12 results for $e^P(x) \equiv (4/9) e^{uV}(x) - (1/9) e^{dV}(x)$ at $Q^2 = 1\, {\rm GeV}^2$ from Ref.~\cite{Courtoy:2022kca}.  
(b) A polynomial fit to the same CLAS12 data.}
\end{figure*}

With the approximations of Ref.~\cite{Seng:2018wwp} and re-derived in Sec.~\ref{sec:twist3}, these scalings also suggest that $\mathcal M_3[e^{u}] + \mathcal M_3[e^{d}] \sim O(N_c)$ and $\mathcal M_3[e^{u}] - \mathcal M_3[e^{d}] \sim O(N_c^0)$.
This can be compared with the scaling of the twist-3 distribution $e^q(x)$ found in the chiral quark soliton model \cite{Schweitzer:2003uy, Cebulla:2007ej, Efremov:2002qh}, which is equivalent to the nonrelativistic quark model in the limit $N_c \to \infty$ \cite{Manohar:1984ys}.
There it was found that $e^{(u+d)} \sim O(N_c^2)$ and $e^{(u-d)} \sim O(N_c)$.
Since the scaling of the full distribution should also be respected by the moments, our relative scalings of the isoscalar and isovector components are in agreement, namely, the isovector component is $1/N_c$ suppressed relative to the isoscalar component.
The differences in the overall scaling we find compared to those of Refs.~\cite{Schweitzer:2003uy, Cebulla:2007ej, Efremov:2002qh} can be traced back to a factor of $m_N \sim O(N_c)$ arising from their use of relativistic normalization of the nucleon states.

\section{Results}
\label{sec:results}
We now estimate the NMEs of the cMDMs in Eqs. (\ref{e:main1}) and (\ref{e:main2}) by taking the constraints on $B^q$ from the spin-flavor expansion and obtaining $e(x)$ from the recent experimental data and model calculations. Having the values of the NMEs, we compare the contributions to the pion-nucleon couplings with those from the quark sigma terms. 
\subsection{cMDMs with the CLAS12 data}

We first estimate the cMDMs by obtaining $e(x)$ from fitting the experimental data, which were recently reported in Ref.~\cite{Courtoy:2022kca} through SIDIS dihadron production. It should be noted that the available data of $e(x)$ corresponds only to the proton flavor combination, given by
\begin{equation}
    e^P(x) \equiv \frac{4}{9} e^{uV}(x) - \frac{1}{9} e^{dV}(x),
\end{equation} 
which accounts solely for the valence quark distributions, defined as $e^{qV}(x) = e^{q} (x) - e^{\overline{q}}(x)$. This implies that extracting the individual distributions $e^{uV}(x)$ and $e^{dV}(x)$ requires additional independent measurements or theoretical constraints. If another independent flavor-sensitive measurement were available---such as a similar extraction from neutron data---it would be possible to solve for $e^{uV}(x)$ and $e^{dV}(x)$ separately. Due to this limitation, to obtain the flavor decomposition, we adopt the theoretical assumption $e^{uV}(x) \approx e^{dV}(x)$, as suggested by the large-$N_c$ approximation. That is, at leading order in the $1/N_c$ expansion, the proton treats up and down valence quarks as nearly identical in their distributions. Applying this approximation to the proton flavor combination, we obtain:
\begin{align}
e^P(x) &= \frac{4}{9} e^V (x)- \frac{1}{9} e^V(x) = \frac{1}{3} e^V(x),
\end{align}
where we define $e^V(x) \equiv e^{uV}(x) \approx e^{dV}(x)$. This relation implies that the extracted $e^P(x)$ directly corresponds to one-third of the common valence distribution. Furthermore, we neglect the antiquark distributions $e^{\overline{u}}(x)=0=e^{\overline{d}}(x)$ because the available data provides only the valence quark combination, whereas we are ultimately interested in the moment:
\begin{align}
\int^{1}_{-1} dx x^{2} e^{u}(x) & = \int^1_0 dx x^{2} (e^{u}(x)+e^{\overline{u}}(x)) \nonumber \\
& \xrightarrow[\phantom{..}]{e^{\overline{u}}= 0} \int^1_0 dx x^2 e^{u}(x),
\end{align}
and similarly for $d$.

We fit $e^{u}(x)$ (which is now the same as $e^{uV}(x)$) to the CLAS data for the proton flavor combination using the Gaussian fitting form:
\begin{equation}
    e^{u}(x) = A \exp \left( -\frac{(x - x_0)^2}{2\sigma^2} \right),
    \label{e:gauss}
\end{equation}
which is the same fitting function as that used in Ref.~\cite{Seng:2018wwp}. We note that this is a three-parameter fit, involving parameters \( A \), \( x_0 \), and \( \sigma \), performed within a Monte Carlo framework. In this framework, the priors for the parameters are randomly shuffled within predefined bounds before each iteration of the fit, which helps avoid the fitting procedure becoming trapped in local minima. The fit is then performed by minimizing the standard (uncorrelated) chi-squared function. Experimental uncertainties are incorporated into the fit by generating synthetic data, which simulates possible outcomes based on the experimental measurements and their uncertainties.
Figure~\ref{fig:figure1} presents the fitting result, which corresponds to the mean values of the fitted curves, with the uncertainty band representing the \( 1\sigma \) standard deviation of the fitted results. Our mean chi-squared value is approximately 1.6, 
after excluding outliers identified using the interquartile range method.

We next test a polynomial functional form of the type: 
\begin{align}
    e^{uV}(x) = N x^\alpha (1-x)^\beta,
\end{align}
and fit all three parameters: \( N \), \( \alpha \), and \( \beta \). However, we encounter difficulties fitting all of them simultaneously. To circumvent this issue, we scan \( \beta \) in the range \( 1 \leq \beta \leq 5 \) and fix its value while fitting the other two parameters using the Monte Carlo method described earlier.  We find that, in terms of minimizing chi-squared, larger values of \( \beta \) yield better results. However, the choice of \( \beta \) is largely dictated by the behavior of the data at large \( x \), where the statistics are sparse, so this should be interpreted with caution. Figure \ref{fig:figure2} shows the corresponding fitting result using this polynomial functional form. Setting \( \beta = 5 \), we obtain a mean chi-squared value (after removing outliers as before) of approximately 3, which is worse than the Gaussian fit.

From these fits, we obtain:
\begin{itemize}
    \item \textbf{Gaussian fit}:
    \[ {\cal M}_3[e^u] + {\cal M}_3[e^d] = 0.2078 \pm 0.1356, \]
    \item \textbf{Polynomial fit}:
    \[ {\cal M}_3[e^u] + {\cal M}_3[e^d] = 0.2606 \pm 0.1750. \]
\end{itemize}  
It is found that the values for the third moment of $e(x)$ from both the gaussian and polynomial fis, within uncertainties, are consistent with the findings of Ref.~\cite{Seng:2018wwp} that gives $0.03<{\cal M}_3[e^u] + {\cal M}_3[e^d]<0.13$ at $Q^2=1.5~{\rm GeV}^2$. Note that in the current analysis, we find that ${\cal M}_3[e^u] - {\cal M}_3[e^d] = 0$, which follows from the assumption $e^u(x) \approx e^d(x)$ based on the large-$N_c$ approximation.

Using the above results of ${\cal M}_3[e^q]$, we now estimate the NMEs of the isoscalar and isovector cMDMs by applying the key constraints on $B^q$ derived from the spin-flavor analysis:
\begin{align}
    B^u + B^d = \pm N_c (1+\delta), \quad B^u - B^d = \pm 1 (1+\delta) \, ,
    \label{const_Bq}
\end{align}  
where $\delta$ accounts for the uncertainty from the truncation of the spin-flavor expansion at $\mathcal{O}(1/N_c^2)$  by incorporating random fluctuations with a standard deviation of approximately 0.3\footnote{In principle, \( \delta \sim \frac{1}{N_c^2} \sim 0.1 \). However, we take a more conservative estimate of \( \delta = 0.3 \), as each term in the spin-flavor expansion also carries an undetermined coefficient that is naively of \( \mathcal{O}(1) \).}.  
The numerical results for $\sigma^{0/3}_C$ are presented in Table~\ref{tab1} for the Gaussian fit and Table~\ref{tab2} for the Polynomial fit.
The $\sigma^0_C$ results between the two fitting forms are broadly consistent. 
Note that \( \sigma^3_C \) is independent of the type of fit, as in the large-\( N_c \) approximation, the difference in the moments of \( e(x) \) for \( u \) and \( d \) quarks cancels out (as mentioned before), which entirely determines \( \sigma^3_C \) by the \( B^{u/d} \) term alone. 
Notably, the cMDM values obtained in our analysis differ from those in Ref.~\cite{Seng:2018wwp}. As shown, our results yield $\sigma^{0/3}_C \sim {\cal O}(1)~{\rm GeV}^2$, whereas the previous study reports $\sigma^0_C = (0.08 - 0.34)~{\rm GeV}^2$ and $\sigma^3_C = 0~{\rm GeV}^2$.  
Our findings and potential uncertainties in the current analysis are summarized below:
\begin{itemize}
\item The dominant contribution to the cMDM arises from the nonzero nucleon form factors $B^q$, leading to larger NME values compared to those reported in the previous study \cite{Seng:2018wwp}, which neglected the $B^q$ terms. In contrast, the contribution from the third moment of $e(x)$ is relatively small (see, however, the caveats discussed below), and the values are somewhat smaller than the expectation from the spin-flavor symmetry, ${\cal M}_3(e^u)+{\cal M}_3(e^d)\sim O(N_c)$.

\item The twist-3 PDF $e(x)$ at large $x$ contributes most significantly to the (third) moment. However, the experimental data available in this tail region are sparse and less constrained, meaning that even small uncertainties can have a significant impact on the computed moment.

\item We emphasize that the effects of antiquarks---presumably smaller---have been neglected in this first assessment, given the large experimental and theoretical uncertainties present in our analysis. 
\end{itemize}

\renewcommand{\arraystretch}{1.5} 
\begin{longtable}[t]{|c|c|}
\caption{Values of \( \sigma^{0/3}_C \) for a Gaussian fit at $Q^2 = 1\, {\rm GeV}^2$.} \\
\hline
\textbf{Parameter Set} & \textbf{cMDM values} \\
\hline
\endfirsthead
\hline
\endfoot
\hline
\begin{tabular}{rl}
    \( B^u + B^d \) & \( = + N_c (1 + \delta) \) \\[6pt]
    \( B^u - B^d \) & \( = + 1 (1+\delta) \) 
\end{tabular} &
\begin{tabular}{rl}
    \( \sigma^0_C \) & \( = + 4.5833 \pm 0.3588 \, {\rm GeV}^2 \) \\[6pt]
    \( \sigma^3_C \) & \( = -2.6891 \pm 0 \, {\rm GeV}^2 \)
\end{tabular} \\
\hline
\begin{tabular}{rl}
    \( B^u + B^d \) & \( = - N_c (1 + \delta) \) \\[6pt]
    \( B^u - B^d \) & \( = - 1 (1+\delta) \) 
\end{tabular} &
\begin{tabular}{rl}
    \( \sigma^0_C \) & \( = -3.4841 \pm 0.3588 \, {\rm GeV}^2 \) \\[6pt]
    \( \sigma^3_C \) & \( = + 2.6891 \pm 0 \, {\rm GeV}^2 \)
\end{tabular} 
\label{tab1}
\end{longtable}
\renewcommand{\arraystretch}{1.5} 
\begin{longtable}[h]{|c|c|}
\caption{Values of \( \sigma^{0/3}_C \) for a Polynomial fit at $Q^2 = 1\, {\rm GeV}^2$.} \\
\hline
\textbf{Parameter Set} & \textbf{cMDM values} \\
\hline
\endfirsthead
\hline
\endfoot
\hline
\begin{tabular}{rl}
    \( B^u + B^d \) & \( = + N_c (1 + \delta) \) \\[6pt]
    \( B^u - B^d \) & \( = + 1 (1+\delta) \) 
\end{tabular} &
\begin{tabular}{rl}
    \( \sigma^0_C \) & \( = + 4.7230 \pm 0.4629 \, {\rm GeV}^2 \) \\[6pt]
    \( \sigma^3_C \) & \( = -2.6891 \pm 0 \, {\rm GeV}^2 \)
\end{tabular} \\
\hline
\begin{tabular}{rl}
    \( B^u + B^d \) & \( = - N_c (1 + \delta) \) \\[6pt]
    \( B^u - B^d \) & \( = - 1 (1+\delta) \) 
\end{tabular} &
\begin{tabular}{rl}
    \( \sigma^0_C \) & \( = -3.3443 \pm 0.4629 \, {\rm GeV}^2 \) \\[6pt]
    \( \sigma^3_C \) & \( = + 2.6891 \pm 0 \, {\rm GeV}^2 \)
\end{tabular}
\label{tab2}
\end{longtable}

\subsection{cMDMs with model calculations}
The impact of incorporating contributions from a full flavor decomposition of $e(x)$, without relying on the large-$N_c$ approximation, as well as the contributions from large-$x$ (which we are currently addressing through extrapolations based on fits to experimental extractions of $e(x)$ over a limited $x$-region), and of including antiquarks, can be examined through model calculations. 

\begin{figure}[t]
    \centering
    \includegraphics[width=0.5\textwidth]{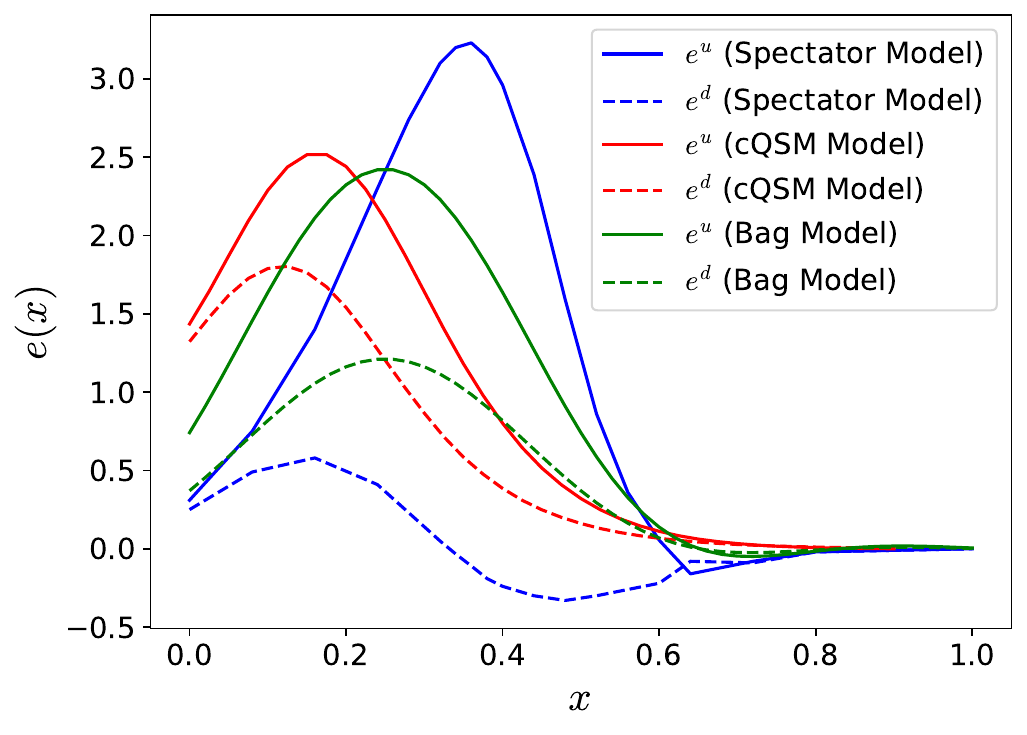} 
    \caption{Comparison of \( e^u(x) \) and \( e^d(x) \) from different models: the spectator model, chiral quark soliton model (cQSM), and bag model. The spectator and bag model results are at a scale of 500 MeV, while the cQSM model result is at 600 MeV. Solid lines represent \( e^u(x) \), while dashed lines represent \( e^d(x)\).}
    \label{fig:model}
\end{figure}

In order to assess these theoretical uncertainties involved in our analysis, we now examine the third moments of $e(x)$ employing theoretical calculations in three models\footnote{We are very grateful to P. Schweitzer for providing the tables for $e(x)$ from the spectator model~\cite{Jakob:1997wg}, the chiral quark soliton model~\cite{Schweitzer:2003uy}, and the bag model~\cite{Jaffe:1991ra}. These resources were instrumental in facilitating our calculations of their contributions to the cMDM.} (see Fig.~\ref{fig:model}). The numerical results of the moment are:

\begin{itemize}
    \item \textbf{Spectator model}~\cite{Jakob:1997wg}: 
    \begin{align*}
        {\cal M}_3[e^u] + {\cal M}_3[e^d] &= 0.0806 , \\
        {\cal M}_3[e^u] - {\cal M}_3[e^d] &= 0.1135 . 
    \end{align*}
    
    \item \textbf{Chiral quark soliton model (cQSM)}~\cite{Schweitzer:2003uy, Cebulla:2007ej}: 
    \begin{align*}
        {\cal M}_3[e^u] + {\cal M}_3[e^d] &= 0.0801 , \\
        {\cal M}_3[e^u] - {\cal M}_3[e^d] &= 0.0220 .
    \end{align*}
    
    \item \textbf{Bag model}~\cite{Jaffe:1991ra}: 
    \begin{align*}
        {\cal M}_3[e^u] + {\cal M}_3[e^d] &= 0.1100 , \\
        {\cal M}_3[e^u] - {\cal M}_3[e^d] &= 0.0367 .
    \end{align*}
\end{itemize}
These moments (not the \(x\)-dependent PDFs themselves) have been evolved from the initial scales at which they were available in models (500 MeV for spectator and bag models, and 600 MeV for cSQM) to \(Q^2 = 1 \, \text{GeV}^2\). This evolution facilitates a direct comparison with the moments obtained from the CLAS experimental data at \(Q^2 = 1 \, \text{GeV}^2\). The evolution follows the equation~\cite{Koike:1996bs}:
\begin{align}
    {\cal M}_3 [e^q](\mu) = \left(\frac{\alpha_s(\mu)}{\alpha(\mu_0)}\right)^{\frac{6.11}{b}} {\cal M}_3 [e^q] (\mu_0),
\end{align}
where \( b = \frac{11 N_c - 2 N_f}{3} \), which is part of an improved evolution formula that includes \(1/{N_c^2}\) corrections in the chiral limit\footnote{This point regarding evolution is subtle, as there is a priori no reason to expect that models should follow the same evolution as that derived from QCD. Therefore, this should be interpreted with caution.
}. \(\alpha_s(\mu)\) and \(\alpha_s(\mu_0)\) represent the strong coupling constants at scales \(\mu\) and \(\mu_0\), respectively.

Overall, the quantity \( {\cal M}_3[e^u] + {\cal M}_3[e^d] \) from the model calculations is consistent with our experimental findings, within uncertainties.
Note that in these models, we have now been able to estimate \( {\cal M}_3[e^u] - {\cal M}_3[e^d] \), which was not possible with the experimental data due to the necessity of using the large-\( N_c \) approximation. It is found that these values are smaller than \( {\cal M}_3[e^u] + {\cal M}_3[e^d] \), except for the one from the spectator model. This exception arises because the down-quark distributions change sign in this model; see Fig.~\ref{fig:model}.

A few comments regarding the antiquark contribution are in order.
In the spectator model, antiquark contributions vanish at leading order—the order at which quark contributions are evaluated here. Incorporating antiquarks, which correspond to higher Fock state components, requires going beyond the leading-order approximation~\cite{Lorce:2014hxa}. 
Since we do not extend the analysis here, the assumption that antiquark contributions are zero remains exact within the scope of this model at the order considered for calculating the moment of $e(x)$. 
However, in the bag model, antiquarks are not zero, although they are unphysical. 
On the other hand, in the cQSM, antiquarks are physical, making this model suitable for estimation. 
Specifically, we estimate from cQSM that the antiquark contribution to the third moment, \( {\cal M}_3[e^{\overline{u}}] + {\cal M}_3[e^{\overline{d}}] \), which is not shown in Fig.~\ref{fig:model}, is 0.0066. Given that the quark contribution is 0.0801, we infer that the antiquark contribution constitutes less than 10$\%$ of the total (quark + antiquark) contribution, thereby justifying its omission in the experimental data analysis. 
The same argument applies to \( {\cal M}_3[e^{\overline{u}}] - {\cal M}_3[e^{\overline{d}}] \). 

With the moments from the model calculations of $e(x)$ as well as the constraints on $B^q$ in Eq. \ref{const_Bq}, we estimate the NMEs of the cMDM. The results for \( \sigma^0_C \) and \( \sigma^3_C \) are presented in Tables~\ref{tab3} and~\ref{tab4}. The values of \( \sigma^0_C \) are generally consistent with those obtained from considering experimental data. For \( \sigma^3_C \), we find that the spectator model produces a slightly off result, which can be attributed to the sign change in the \( d \)-quark distribution within this model (as mentioned before). Nevertheless, the results for $\sigma^3_C$ from other models are in good agreement with our findings from experimental data.

In conclusion, through these model estimates, we have sought to justify and provide an estimate of the theoretical uncertainties underlying our analysis of the experimental data. We reiterate that, overall, the range of cMDM values from experimental data agrees with our quark model calculations of $e(x)$. These estimates are likely valid within a typical quark model uncertainty of $20\%-30\%$, though future studies will be necessary to draw more definitive conclusions. 
With these two approaches to the extraction of the twist-3 PDF $e(x)$, we conclude that the nucleon form factor of \( B^q \) dominates the cMDMs, while the contribution from the third moment of \( e(x) \) is at most ${\cal O} (10)\%$.

\renewcommand{\arraystretch}{1.5} 
\begin{longtable}[ht]{|c|c|}
\caption{Values of \( \sigma^{0}_C \) from models at $Q^2 = 1\, {\rm GeV}^2$.} \\
\hline
\textbf{Parameter Set} & \textbf{cMDM values} \\
\hline
\endfirsthead
\hline
\endfoot
\hline
\begin{tabular}{rl}
    $ B^u + B^d $ & $ = + N_c (1 + \delta) $ 
\end{tabular} &
\begin{tabular}{rl}
    $ \sigma^0_C \big |_{\rm spectator} $ & $ = + 4.2469  \, {\rm GeV}^2 $ \\[6pt]
    $ \sigma^0_C \big |_{\rm cQSM}$ & $ = + 4.2456  \, {\rm GeV}^2 $ \\[6pt]
    $ \sigma^0_C \big |_{\rm bag}$ & $ = + 4.3247  \, {\rm GeV}^2 $ \\[6pt]
\end{tabular} \\
\hline
\begin{tabular}{rl}
    $ B^u + B^d $ & $ = - N_c (1 + \delta) $ 
\end{tabular} &
\begin{tabular}{rl}
    $ \sigma^0_C \big |_{\rm spectator} $ & $ = -3.8205  \, {\rm GeV}^2 $ \\[6pt]
    $ \sigma^0_C \big |_{\rm cQSM}$ & $ = -3.8218  \, {\rm GeV}^2 $ \\[6pt]
    $ \sigma^0_C \big |_{\rm bag}$ & $ = -3.7427  \, {\rm GeV}^2 $ \\[6pt]
\end{tabular} 
\label{tab3}
\end{longtable}
\renewcommand{\arraystretch}{1.5} 
\begin{longtable}[c]{|c|c|}
\caption{Values of \( \sigma^{3}_C \) from models at $Q^2 = 1\, {\rm GeV}^2$.} \\
\hline
\textbf{Parameter Set} & \textbf{cMDM values} \\
\hline
\endfirsthead
\hline
\endfoot
\hline
\begin{tabular}{rl}
    \( B^u - B^d \) & \( = +1 (1+\delta) \) 
\end{tabular} &
\begin{tabular}{rl}
    $ \sigma^3_C \big |_{\rm spectator}$ & $ = -3.2896  \, {\rm GeV}^2 $ \\[6pt]
    $ \sigma^3_C \big |_{\rm cQSM}$ & $ = -2.8055  \, {\rm GeV}^2 $ \\[6pt]
    $ \sigma^3_C \big |_{\rm bag}$ & $ = -2.8833  \, {\rm GeV}^2 $ \\[6pt]
\end{tabular} \\
\hline
\begin{tabular}{rl}
    $ B^u - B^d $ & $ = - 1(1+\delta) $ 
\end{tabular} &
\begin{tabular}{rl}
    $ \sigma^3_C  \big |_{\rm spectator}$ & $ = + 2.0887
   \, {\rm GeV}^2 $ \\[6pt]
    $ \sigma^3_C  \big |_{\rm cQSM}$ & $ = + 2.5727  \, {\rm GeV}^2 $ \\[6pt]
    $ \sigma^3_C  \big |_{\rm bag}$ & $ = + 2.4950 \, {\rm GeV}^2 $ \\[6pt]
\end{tabular} 
\label{tab4}
\end{longtable}

\subsection{Contributions to the pion-nucleon couplings}
Having the results of $\sigma^{0/3}_C$ discussed in the previous section, we now compare their contributions to the pion-nucleon couplings in Eqs.~(\ref{eq:pN_0}) and (\ref{eq:pN_1}) with those from the quark sigma terms $\sigma^{0/3}$. By simply averaging the results from Tables~\ref{tab1}, \ref{tab2}, \ref{tab3}, and \ref{tab4}, we obtain  
\begin{align}
\sigma^{3}_C=\begin{cases}
        +2.5\\
        -2.9\\
    \end{cases}
    \hspace{-0.2cm}{\rm GeV}^2,\hspace{0.5cm}
\frac{r\sigma^3}{\bar{m}\varepsilon}=0.87~{\rm GeV}^2,
\end{align}
for the isovector coupling $\bar{g}_0$, and
\begin{align}
    \sigma^0_C=
    \begin{cases}
        +4.4\\
        -3.6\\
    \end{cases}
    \hspace{-0.2cm}{\rm GeV}^2,\hspace{0.5cm}
    \frac{r\sigma^0}{\bar{m}}=5.9~{\rm GeV}^2,
\end{align}
for the isoscalar coupling $\bar{g}_1$. Overall, our results imply that the contributions from the cMDMs can be the same order of magnitude as those from the quark sigma terms. Having these results, we obtain
\begin{align}
    \bar{g}_0&=\left(\tilde{d}_u+\tilde{d}_d \right)\times
    \begin{cases}
        +1.7\\
        -1.0
    \end{cases}\hspace{-0.2cm}{\rm GeV}^2,\\
    \bar{g}_1&=-\left(\tilde{d}_u-\tilde{d}_d \right)\times
    \begin{cases}
        -1.5\\
        -9.6
    \end{cases}\hspace{-0.2cm}{\rm GeV}^2,
\end{align}
leading to $\left|{\bar{g}_1}/{\bar{g}_0}\right|\gtrsim 1$ modulo the quark cEDM couplings. Note that the former study concludes that $\bar{g}_1\gg \bar{g}_0$ resulting from  the quark sigma terms \cite{Seng:2018wwp}. Finally, comparing with the preferred range obtained by the QCD Sum Rule \cite{Pospelov:2001ys} (see Table 1 therein),
\begin{align}
    \bar{g}_0&=\left(\tilde{d}_u+\tilde{d}_d \right)\times
    \begin{cases}
        +0.2\\
        -0.5
    \end{cases}\hspace{-0.2cm}{\rm GeV}^2,\\
    \bar{g}_1&=\left(\tilde{d}_u-\tilde{d}_d \right)\times
    \begin{cases}
        -0.4\\
        -2.2
    \end{cases}\hspace{-0.2cm}{\rm GeV}^2 .
\end{align}
One can see that our results are larger by a factor of 4 or 5.

We emphasize that the results presented here should be interpreted with caution; the spin-flavor analysis allows us to constrain the unknown nucleon form factors \( B^q \), which was considered to be subleading in previous work. However, it only provides a rough upper bound on the order or $O(N_c)$ rather than definite numerical values. Together with the extraction of the third moment of \( e(x) \), we can infer that the contributions from the form factors to the cMDMs are relatively large. Nevertheless, the current study remains far from a precise determination of \( \sigma^{0,3}_C \). We hope that our work motivates further efforts to evaluate the NMEs through various approaches, particularly direct lattice QCD calculations of the matrix elements~\cite{Walker-Loud}.

\section{Conclusions}
\label{sec:conclusion}
We have estimated the nucleon matrix elements of quark chromo-magnetic dipole moments relevant for CP-odd pion-nucleon interactions induced by quark chromo-electric dipole moments. The NMEs can be expressed by two parameters: nucleon form factors $B^q~(q=u,d)$ and the third moment of the twist-3 parton distribution function (PDF) $e(x)$. 
Analyzing the cMDM operators with the spin-flavor expansion from the large-$N_c$ limit of QCD \cite{Dashen:1994qi}, we point out that the nucleon form factors, which are assumed to be negligible in the former study \cite{Seng:2018wwp}, can be ${\cal O}(N_c)$ for the isoscalar combination of $B^q$ and ${\cal O}(N_c^0)$ for the isovector one. 
The twist-3 PDF can be extracted through dihadron production in semi-inclusive deep inelastic scattering processes, and the CLAS Collaboration recently updated the experimental extraction \cite{Courtoy:2022kca}. Having the constraints on $B^q$ from the spin-flavor analysis and the latest experimental data on $e(x)$, we find that the nucleon form factors can dominate the NMEs of the cMDM operators over the third moment of the twist 3 PDF. 
This finding is also confirmed by employing model calculations of $e(x)$ from spectator, chiral quark soliton, and bag models.  Our results indicate that the NMEs of the cMDMs can be the same order of magnitude as the other contributions from quark sigma terms to the pion-nucleon couplings. We hope that the argument is ultimately determined by the direct lattice QCD calculations of the NMEs.

\begin{acknowledgments}
We thank T. Bhattacharya, V. Cirigliano, Y. Hatta, D. Pefkou, P. Schweitzer, C-Y. Seng, F. Yuan, and A. Walker-Loud for valuable discussions. K. F. is grateful to N3AS and RIKEN-Berkeley Center at UC Berkeley, where we started a discussion, for the hospitality and support. The work of S.~B., K.~F. and E.~M. was supported by the Laboratory Directed Research and Development program of Los Alamos National Laboratory under project number 20240738PRD1 (S.~B), 20250164ER, 20210190ER, and 20240078DR (K.~F. and E.~M.).
K.~F. and E.~M. were supported by the U.~S. Department of Energy through the Los Alamos National Laboratory. Los Alamos National Laboratory is operated by Triad National Security, LLC, for the National Nuclear Security Administration of the U.S. Department of Energy (Contract No. 89233218CNA000001).
T.~R.~R. is supported through the NSF through cooperative agreement 2020275 and by the DOE Topical Collaboration “Nuclear Theory for New Physics,” award No. DE-SC0023663.
\end{acknowledgments}

\bibliography{refs}
\end{document}